**Compartmental diffusion and microstructural properties of human brain gray and white matter studied with double diffusion encoding magnetic resonance spectroscopy of metabolites and water**

Henrik Lundell*[#1], Chloé Najac*[2], Marjolein Bulk[2], Hermien E. Kan[2], Andrew G. Webb[2], Itamar Ronen[2]

[1]Danish Research Centre for Magnetic Resonance, Centre for Functional and Diagnostic Imaging and Research, Copenhagen University Hospital Hvidovre, Kettegaards Allé 30, 2650 Hvidovre, Denmark

[2]C.J. Gorter Center for High Field MRI, Department of Radiology, Leiden University Medical Center, Albinusdreef 2, 2333 ZA Leiden, The Netherlands

* Authors contributed equally to this work
[#] Corresponding author, e-mail address lundell@drcmr.dk

**Funding:** This project was supported by the European Research Council (ERC) under the European Union's Horizon 2020 research and innovation programme (grant agreement No 804746), by the Leiden University Fund (W-19356-2-32) and LEaDing Fellows COFUND programme. M.B. was supported by grant from ZonMw program Innovative Medical Devices Initiative, project Imaging Dementia: Brain Matters (104003005)
**Competing interests:** None

# Abstract

Double diffusion encoding (DDE) magnetic resonance measurements of the water signal offers a unique ability to separate the effect of microscopic anisotropic diffusion in structural units of tissue from the overall macroscopic orientational distribution of cells. However, the specificity in detected microscopic anisotropy is limited as the signal is averaged over different cell types and across tissue compartments.

Performing side-by-side metabolite DDE spectroscopy (DDES) and water DDES in which a wide range of *b*-values is used to gradually eliminate the extracellular contribution provides complementary measures from which intracellular and extracellular microscopic fractional anisotropies (µFA) and diffusivities can be estimated. Metabolites are largely confined to the intracellular space and therefore provide a benchmark for intracellular diffusivity of specific cell types. Here, we aimed to estimate tissue- and compartment-specific human brain microstructure by combining water and metabolites DDES experiments.

We performed DDES in human subjects in two brain regions that contain widely different amounts of white matter (WM) and gray matter (GM): parietal white matter (PWM) and occipital gray matter (OGM) on a 7 T MRI scanner. Results of the metabolite DDES experiments in both PWM and OGM suggest a highly anisotropic intracellular space within neurons and glia, with the possible exception of gray matter glia. Tortuosity values in the cytoplasm for water and tNAA, obtained with correlation analysis of microscopic parallel diffusivity with respect to GM/WM tissue fraction in the volume of interest, are remarkably similar for both molecules, while exhibiting a clear difference between gray and white matter, suggesting a more crowded cytoplasm and more complex cytomorphology of neuronal cell bodies and dendrites in GM than those found in long-range axons in WM.

## Introduction

Diffusion-weighted MRI (DW-MRI) sensitizes the signal to molecular displacement on length scales comparable to cell morphological features, making DW-MRI a sensitive tool for tissue characterization on a microscopic scale[1]. Inferring specific cytomorphological properties from DW-MRI is not straightforward, as diffusion measurements reflect the average properties across a large volume compared to cellular dimensions. This average reflects properties that range from the microscopic shape and size of restricting geometries of different cell types to the heterogeneous organization over the entire voxel. Some microstructural information may be retrieved from modeling conventional diffusion data, thereby accounting for the heterogeneity across the imaging voxel. This, however, heavily relies on model assumptions that are difficult to validate with the ambiguous and sparse histological data available[2,3].

A different approach is to increase specificity to the underlying cytomorphological features at the data acquisition stage. Several methods have been suggested to separate anisotropic diffusion on the *microscopic* scale, reflecting the presence of thin fibers such as axons and astrocytic processes, from *macroscopic* anisotropy, which reflects co-alignment of fibers across the acquisition volume. These methods include double diffusion encoding (DDE) experiments[4,5], related tensor valued encoding methods in multiple directions with tailored gradient waveforms[6], and analysis of the non-monoexponential attenuation of disordered samples or the so-called "powder average" of diffusion measurements in multiple directions with respect to the *b* value in conventional diffusion experiments[7–12]. Microscopic anisotropy (µFA) and derived microscopic axial ($D_{//}$) and transverse ($D_\perp$) diffusivities can be calculated from these measurements, and these reflect the diffusion properties within cytomorphological units on the length scale of the diffusion process, regardless of their mutual orientation on larger length scales (Figure 1A)[13–15]. Compartmental and cell-type selectivity, and in some cases specificity, can be obtained with diffusion-weighted magnetic resonance spectroscopy (DW-MRS)[16,17], where the diffusing microstructural probes are intracellular metabolites. DW-MRS is thus not only specific to the intracellular space but also allows studying cytomorphology of different cell populations such as neurons and glia across species (figure 1B)[18]. The compartmental specificity of DW-MRS extends beyond obtaining standard diffusional metrics such as metabolite apparent diffusion coefficients (ADC). The combination of DDE and DW-MRS (DDES) [19–22] can yield cell-type-specific microscopic diffusion metrics of the intracellular space, and contribute to a more detailed characterization of cytomorphology in neural tissue.

The morphological properties of the intracellular and extracellular spaces are not independent of one another, especially in white matter (WM), where the dominant structural unit is long-range axons in WM tracts. The relatively compact packing of axons in WM suggests the possibility of microscopic anisotropy also in the extracellular space[23,24]. Earlier work interpreting conventional DW-MRI with biophysical models suggests that in WM the diffusivity in the extracellular space ($D_{//}$(extracellular water) ~ 2 µm²/ms and $D_⊥$(extracellular water) ~ 0.5 µm²/ms for an overall D(extracellular water) ~ 1 µm²/ms) is higher than in the intracellular space ($D_{//}$(intracellular water) ~ 2 µm²/ms and $D_⊥$(intracellular water) ~ 0 µm²/ms for an overall D(intracellular water) ~ 0.67 µm²/ms)[25–28]. This suggests that diffusion weighting modulates, and at high *b* value effectively eliminates, the contribution from the extracellular space. This has inspired the notion of using large diffusion weightings as a filter to target measurements of intra-axonal water[12,29–31] (figure 1C). For a quantitative assessment of the elimination of extracellular water in our experiments, see figure S1 in the appendix.

The microstructural characterization of both the intracellular and the extracellular spaces in gray matter (GM) has been less investigated. The microstructural heterogeneity of human GM, in addition to the lack of evidence of macroscopic anisotropy except for that found in neonates[32] and some cortical regions[33], has discouraged researchers from further exploring the microstructural properties of GM in detail with DW-MRI. Pioneering work with DDE demonstrated the presence of microscopic anisotropy in GM of excised pig spinal cord and brain samples[34–36]. Only recently have attempts been made to characterize microscopic anisotropy in human GM *in vivo*, either with tensor valued encoding techniques[37] or with DDE[38].

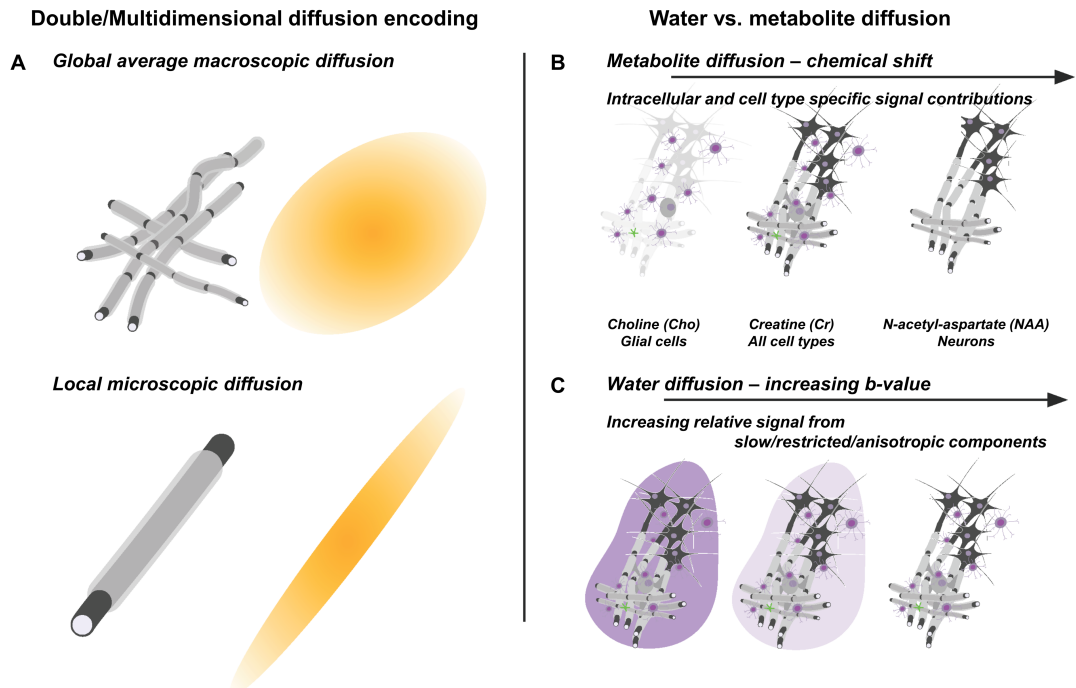

*Figure 1*: Approaches used in this study to provide compartment-specific readouts of diffusivity: (A) Double diffusion encoding (DDE) enables the decoupling of global effects of fibers across the acquisition volume from local microscopic diffusion. (B) Studying metabolite diffusion with DDES provides access to the cell-preferential or specific intracellular microscopic organization. (C) Measuring water diffusion with DDES over a large range of b values allows separating CSF, intra- and extracellular properties.

In this study, we present a DDES approach to characterize the microscopic anisotropy and diffusivities of the intracellular and the extracellular spaces in the human brain. The complementarity of the DDES water measurements at low *b* values, which includes contributions from both the intra- and extracellular spaces, with the intracellular-specific DDES measurements of metabolites and water at high *b* values results in a set of unique quantitative insights on the morphology of the intracellular space in gray and white matter. In addition, qualitative assessments of the microstructural characteristics of the extracellular space are also derived.

# Materials and methods

## Human subjects

A total of 20 healthy volunteers (age 33±12 years, 11 females) participated in the study, with 2 volunteers participating twice for acquisitions in two different regions (see description below). The study adhered to the guidelines of the Leiden University Medical Center Institutional Review Board (The Netherlands). Informed consent was obtained from all subjects prior to the session.

## MRI scanner/hardware

All experiments were performed on a 7T Philips Achieva whole-body MRI scanner (Philips Healthcare, Best, The Netherlands) equipped with a volume transmit/32-channel receive head coil (Nova Medical, Wilmington MA, USA) and gradient coils with a maximum gradient strength of 40 mT/m and a slew rate of 200 T/m/s. A high permittivity dielectric pad (suspension of barium titanate in $D_2O$) was used to maximize the transmit magnetic field ($B_1^+$) homogeneity and efficiency in the parietal and occipital regions as previously described[39,40].

## MRI/DW-MRS data acquisition

The acquisition time for the entire protocol averaged approximately 55 minutes and comprised the following scans.

### Anatomical Images

A short survey scan and a sensitivity encoding (SENSE) reference scan followed by a 3D $T_1$-weighted gradient-echo acquisition were conducted to allow for the planning of the volumes-of-interest (VOIs) for the DDES experiments (figure 2). Imaging parameters for the 3D-$T_1$ weighted scan were: field-of-view (a-p, f-h, r-l): 246x246x174 $mm^3$, in-plane resolution: 1x1x1 $mm^3$, repetition time (TR)/echo time (TE): 4.9/2.2 ms, total scan duration: 1:55.

### Water and metabolite DDES

*Pulse sequence*: DDES data were acquired using a DDE-sLASER sequence[41]. The sequence diagram is shown in figure 3A. The following acquisition parameters were used: TE: 185 ms, spectral width: 3000 Hz, number of time-domain points: 1024. Each of the two diffusion weighting modules within our DDES sequence consisted of a double spin-echo with a bipolar diffusion weighting scheme. Using the conventions established for DDE sequences[4] adapted to a bipolar DW scheme: for both DW modules, single gradient lobe duration $δ_1/2 = δ_2/2$ = 15.5 ms; bipolar gap $τ_1 = τ_2$ = 10 ms; gradient separation time $Δ_1 = Δ_2$ = 45 ms; mixing time $t_m$ = 5.3 ms (see figure 3A for definitions of timing parameters).

*Volumes of interest*: We examined two different brain regions: a WM region within the left parietal lobe (PWM) and a GM cortical region within the occipital lobe (OGM). 9 subjects were scanned with a 9 cm$^3$ VOI in the PWM region (3cm (a-p), 2 cm (f-h), 1.5 cm (r-l)); 9 subjects were scanned with a 9 cm$^3$ VOI in the OGM region (3cm (a-p), 1.5 cm (f-h), 2 cm (r-l)). To increase the specificity to GM in the OGM, 4 subjects were scanned with a smaller OGM VOI (sOGM) of 2.5 cm$^3$ (2.5 cm (a-p), 1 cm(f-h), 1cm (r-l)). Two subjects participated twice: one was scanned with PWM and OGM VOIs and another was scanned with OGM and sOGM VOIs. Due to the limited SNR, only water acquisition was possible in the smaller VOI. The positions of the VOIs are illustrated in figure 2. Due to time constraints, metabolites and water DDES scans at all *b* values were also not acquired in all subjects. Table 1 summarizes the sample size for each acquisition performed in PWM and OGM VOIs.

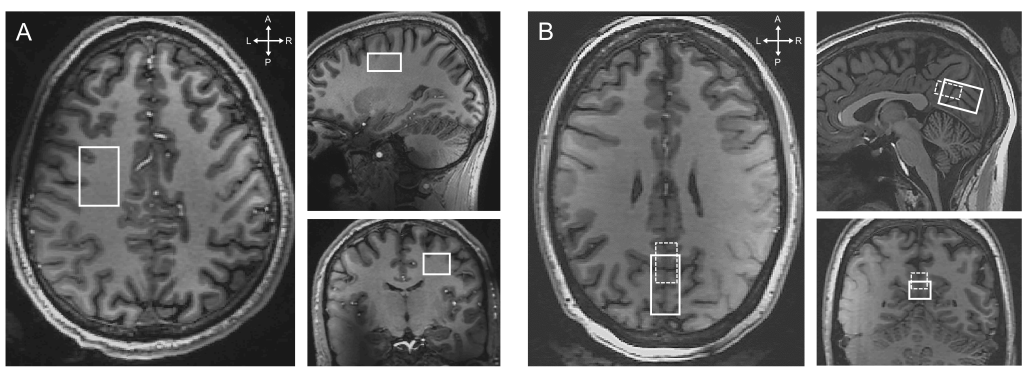

*Figure 2*: Representative placement in the of the DDES VOI in (A) the PWM region and (B) the OGM region. The 9 cc PWM VOIs contained mostly WM (~80%) while the 9cc OGM VOIs contained about 50% GM. The smaller OGM VOI (dashed lines), made to increase the GM fraction within the VOI, contained 68% GM on average.

|  | *Water DDES* | | | *Metabolites DDES* | |
| --- | --- | --- | --- | --- | --- |
|  | *PWM* | *OGM* | *sOGM* | *PWM* | *OGM* |
| *b* = 918 s/mm$^2$ | 6 | 8 | 4 | | |
| *b* = 2066 s/mm$^2$ | 6 | 6 | 4 | | |
| *b* = 4050 s/mm$^2$ | 7 | 6 | 4 | | |
| *b* = 7199 s/mm$^2$ | 8 | 7 | 4 | 6 | 7 |

*Table 1*: Sample size (n) for water and metabolite DDES experiments in PWM and OGM VOIs with the range of b values used.

*Water suppression, $B_0$ shimming, and scan synchronization:* VOI-localized $B_0$ shimming up to second order was performed. To minimize signal fluctuations due to cardiac pulsation, cardiac triggering was achieved using a peripheral pulse unit (trigger delay: 250 ms, TR: 5 cardiac cycles). For metabolite acquisitions, water suppression was achieved using two frequency-selective excitation pulses, each followed by a dephasing gradient.

*DDE acquisition schemes*: The angular DDE encoding was performed in three orthogonal planes spanned by the three vectors [1 1 -0.5], [-0.5 1 1], and [1 -0.5 1]. These directions provide the maximum combined gradient amplitude for three orthogonal directions. For each acquisition, the first encoding direction was fixed and the second encoding was performed in 8 equally spaced directions tracing a full circle, starting with the first encoding direction (see figure 3B). θ refers to the angle between the effective gradient directions of the first and second encoding, adhering to the definition given in Shemesh et al.[4]. Two diffusion gradient amplitudes were used for metabolite acquisitions, resulting in 2 *b* values for the entire diffusion encoding scheme: 0 and 7199 s/mm$^2$. Five diffusion gradient amplitudes, resulting in 5 *b* values (0, 918, 2066, 4050, and 7199 s/mm$^2$) were used for water acquisitions. To compensate for cross-terms between diffusion and imaging/background gradients[42] in the post-processing stage, DW data were acquired alternately with opposite diffusion gradient polarities. Overall, each DW condition was repeated 6 times for metabolites and 2 times for water acquisitions. For metabolite DDES, the total number of acquisitions was $N_{acq}$ = (3 (directions first encoding) x 8 (directions second encoding) x 2 (gradient polarities) + 1 (*b* = 0 condition)) x 6 (number of averages) = 294 acquisitions. With TR of 5 cardiac cycles, the total scan time was about 294 x 5 = 25 minutes. For water DDES, the number of averages was 2, resulting in $N_{acq}$ = 98 and an 8 minute scan.

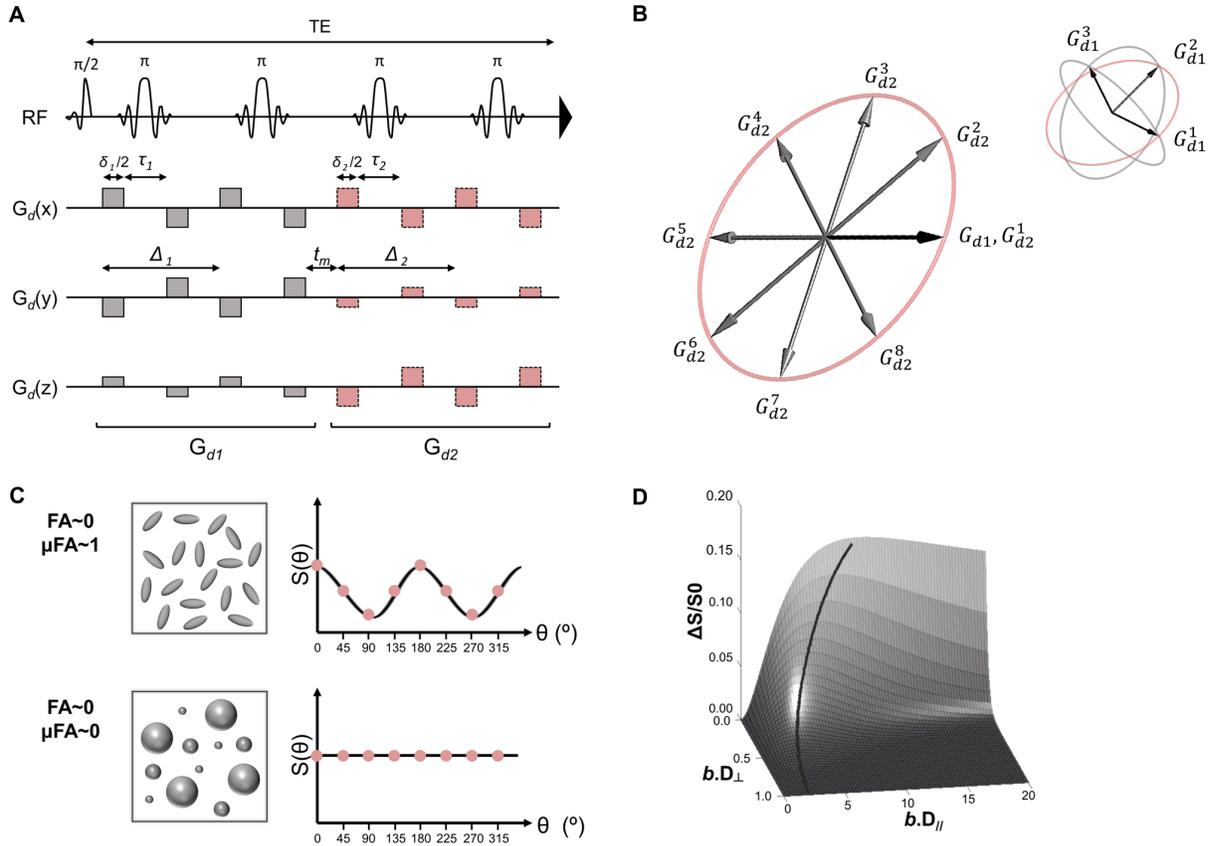

*Figure 3: (A) Schematic drawing of the DDE-sLASER sequence. For simplicity, only diffusion sensitizing gradients are shown. (B) Illustration of diffusion weighting (DW) gradient directions used in our study. The direction of the first DW group ($G_{d1}$) is fixed along one of three orthogonal directions (small panel to the right) while the direction of the second diffusion group ($G_{d2}$) revolves in 8 angular steps (gray vectors) on a circle that includes the first gradient direction (black vector). Gray circles represent the circles traced by $G_{d2}$ for the two remaining directions of $G_{d1}$. (C) Schematic illustration of the modulation of the signal as a function of θ (angle between first and second encoding) expected for ensembles of rotationally disperse anisotropic (top) or isotropic (bottom) diffusion tensors. (D) Illustration of the expected contrast between parallel (θ = 0º) and perpendicular (θ = 90º) encodings for a rotationally uniform distribution of monodisperse diffusion tensors over a range of eigenvalues relative to b. The black curve indicates the maximum contrast at a given transverse diffusivity.*

## Data processing and analysis

*Image processing:* $T_1$-weighted images were segmented into tissue maps for GM, WM and cerebrospinal fluid (CSF) using FSL (Brain extraction Tool[43] and FAST[44] algorithm in the FMRIB Software Library). Each voxel contains a value in the range 0-1 that represents the proportion of each tissue. An in-house Matlab routine (MathWorks, Inc., Ma, USA) was then used to quantify the tissue volumes within each spectroscopic VOI. One subject was excluded from the tissue analysis due to the poor quality of the segmentation.

*DDES Data pre-processing:* Individual spectra were corrected for eddy currents, phase, and frequency drifts using in-house Matlab routines as previously described[45]. Averaged metabolite data sets consisted of 49 individual spectra: one acquired with $b$ = 0 s/mm$^2$ and two sets of 24 (3 ($G_{d1}$) x 8 ($G_{d2}$)) DW spectra at $b$ = 7199 s/mm$^2$, each set acquired with two gradient polarities. DW spectra were subsequently averaged across the three $G_{d1}$ directions to provide an emulated powder average, resulting in two sets of 8 DW spectra. These spectra and the one acquired at $b$ = 0 s/mm$^2$ were subsequently quantified with LCModel[46], resulting in signal amplitudes of total N-acetyl-aspartate (tNAA = N-acetyl aspartate (NAA) + N-acetylaspartylglutamate (NAAG)), total creatine (tCr = creatine (Cr) + phosphocreatine (PCr)) and total choline (tCho = choline (Cho) + phosphocholine (PCho) + glycerophosphocholine (GPC)) for each value of $b$, θ and gradient polarity. The LCModel basis set included a total of 16 metabolites and a control node spacing of the spline function for fitting the baseline (dkntmn) of 0.5. Finally, the geometric mean of the LCModel estimates was calculated for each pair of spectra acquired with the same ($b$, θ) acquired with opposite sign of the gradient polarity, thereby reducing the effect of cross-terms with background and sequence gradients. An evaluation of this effect on phantom data is shown in figure S2 in the appendix. The resulting diffusion-weighted metabolite signals were finally normalized to their respective signal at $b$ = 0 s/mm$^2$. The post-processing scheme is depicted in figure 4.

The water signal was preprocessed similarly, and water spectra acquired with the same diffusion-weighted condition ($b$, θ, gradient polarity) were averaged. The amplitude of the water signal for each averaged spectrum was obtained by integrating the area under the water peak in Matlab. The remaining post-processing procedure for the water data at each of the 4 $b$ values followed the one described above for the metabolites, resulting in a single water signal value for $b$ = 0 s/mm$^2$ and 4 sets of 8 water signal values, for each value of $b$ and θ, respectively. All DW water signals were normalized to the signal at $b$ = 0 s/mm$^2$. A short representation of the raw data is available in tables S2 and S3 in the appendix. The raw data and associated analysis code is available from Itamar Ronen (i.ronen@lumc.nl) upon request and data handling agreement.

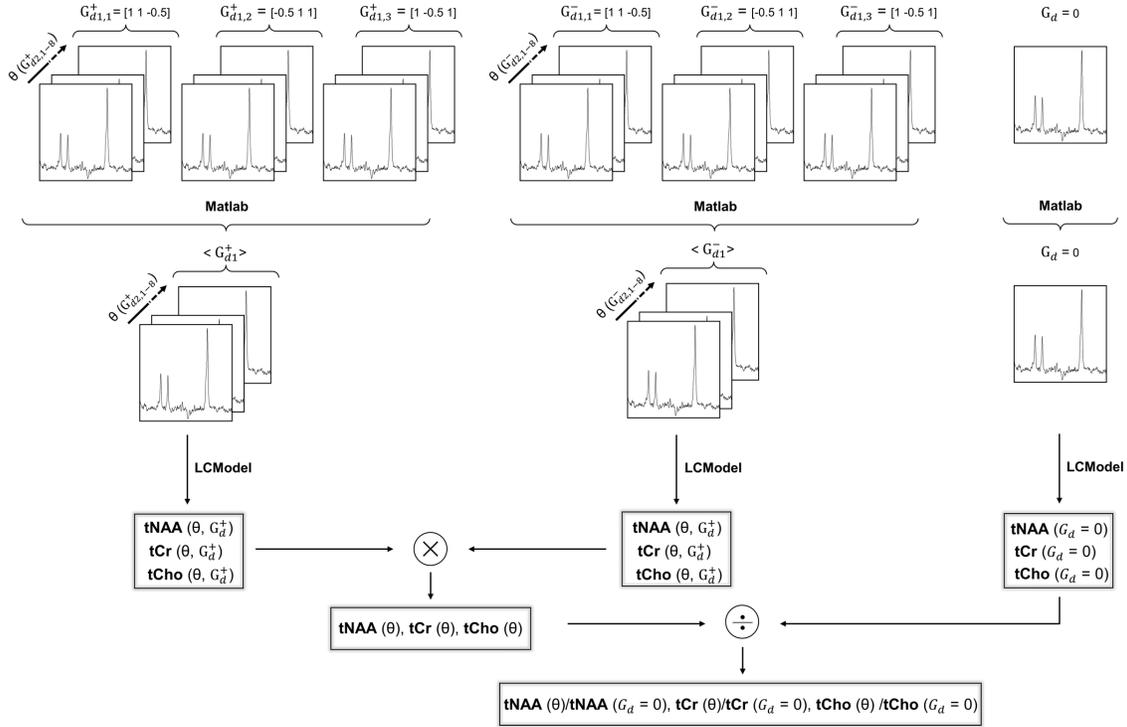

*Figure 4: Illustration of metabolite data pre-processing. Metabolite data sets consisted of individual spectra acquired with (b = 7199 s/mm$^2$) and without (b = 0 s/mm$^2$) diffusion-weighting (DW). DW spectra were alternatingly acquired with opposite gradient polarity ($G_d^+$ and $G_d^-$) and with 24 angular conditions (3 ($G_{d1}$) x 8 ($G_{d2}$)). DW spectra were first powder-averaged across the three $G_{d1}$ values. DW and non-DW spectra were then quantified with LCModel[46] resulting in signal amplitudes of tNAA, tCr, and tCho. The geometric means of estimates with opposite polarity were finally calculated and normalized to their respective non-DW signals.*

## Analyses and interpretation of DDES data

Under the assumption of mono-disperse, axially symmetric diffusion tensors **D** with axial and radial diffusivities $D_{//}$ and $D_\perp$, the *b* value and θ-dependent DDE signal from an ensemble of *N* different orientations described by the rotation matrices $\mathbf{R_j}$ can be calculated as:

$$S(b,\theta) = \frac{S_0}{N} \sum_{j=1}^{N} \exp\left(-\frac{b}{2}\mathbf{e_1 R_j D R_j^T e_1^T}\right) \exp\left(-\frac{b}{2}\left(cos^2\theta \mathbf{e_1 R_j D R_j^T e_1^T} + sin^2\theta \mathbf{e_2 R_j D R_j^T e_2^T}\right)\right)$$

The two orthogonal unit vectors $e_1$ and $e_2$ span the encoding plane. The signal from a powder average was emulated with *N* = 256 uniformly distributed rotations with respect to the axial direction of an axially symmetric diffusion tenor. The two exponentials in the summation reflect the signal attenuation from the first and second diffusion encoding gradient acting on each rotation $\mathbf{R_j}$ of the diffusion tensor **D**. While the signal of a powder average at low *b* values reflects the initial slope or the mean diffusivity of the ensemble, the different angular modulation of anisotropic components with different orientations gives a multiexponential behavior that becomes more apparent at higher *b value*s (shown schematically in figure 3C and simulated for monodisperse diffusion tensors in figure 3D).

For metabolites, the single compartment diffusivity was estimated from the $b = 0$ s/mm$^2$ normalized signal ($S(b,\theta)/S_0$) with the axial ($D_{//}$) and radial ($D\perp$) microscopic diffusivities as fitting parameters. A similar assumption for the water signal is not expected to be valid, as multiple signal components from CSF and different intra- and extracellular environments violate the monodisperse assumption. However, residual restricted and anisotropic components would be expected to dominate the signal at larger *b* values. Here, the water signal was fitted using data from two subsequent *b* values with $S_0$ as an additional fitting parameter. For mixed components with well-separated diffusivities, the fitted $S_0$ at high *b* values thus reflects the residual volume of slow and anisotropic signal components. Some hypothetical considerations for signal contributions from different environments are shown in figure S1 in the appendix. All fitting procedures were performed using non-linear sum of squares error minimization with $D_{//}$ = 1 µm$^2$/ms, $D\perp$ = 0 and $S_0$ = 1 as initial values. The microscopic fractional anisotropy (µFA) was calculated analogous to the fractional anisotropic (FA) from DTI analyses:

$$\mu FA = \sqrt{\frac{(D_{//} - D_\perp)^2}{D_{//}^2 + 2D_\perp^2}}$$

**Statistical analysis**

Results are expressed as mean ± standard deviation. Statistical significance was tested using GraphPad Prism 7 using an unpaired Student's *t* test with unequal variance (GraphPad Software, USA). A threshold of $p < 0.05$ was considered significant, the following symbols where used to indicate the significance: *$p < 0.05$, **$p < 0.01$, ***$p < 0.001$, ****$p \leq 0.0001$. Correlation between intracellular tortuosity and $D_{||}$ with tissue fraction were obtained with linear regression using GraphPad Prism 7.

# Results

## Volume Fraction of GM, WM, and CSF

Table 2 shows the average percentages of WM, GM, and CSF within our two VOIs. The average WM fraction in the PWM VOI is above 80% while the average GM fraction is about 15%. In comparison, the average WM fraction in the OGM VOI is lower than 40% and the average GM fraction is around 50%. To increase the specificity to GM, a subset of data was acquired in a smaller OGM VOI (sOGM), which contained on average around 70% GM and 20% WM (see table S1 in the appendix). GM and WM content were significantly different in OGM VOI compared to PWM VOI ($p < 0.0001$). Finally, the CSF fraction was fairly low in the PWM VOI but was significantly higher in OGM VOI (~14%, $p < 0.006$) as it included the interhemispheric fissure. While the CSF signal does not contribute to the diffusion results at high $b$, it influences the results at the lower $b$, particularly in the OGM VOI.

|  | PWM VOI | OGM VOI | PWM *vs.* OGM |
|---|---|---|---|
| **Gray Matter (GM)** | 15.1±5.1 ‡‡‡‡ | 50.1±7.9 ‡ | **** |
| **White Matter (WM)** | 82.8±5.7 | 36.2±15.1 | **** |
| **Cerebrospinal Fluid (CSF)** | 2.1±1.2 | 13.7±9.1 | ** |

*Table 2*: Volume fraction (%, mean±s.d.) of WM, GM, and CSF in the parietal white matter (PWM) and occipital gray matter (OGM) VOIs. Statistical significance between PWM and OGM VOIs (represented with *) and between GM and WM content (represented with ‡) were evaluated using an unpaired Student's t-test with *, ‡ $p<0.05$, **, ‡‡ $p<0.005$, ***, ‡‡‡ $p<0.001$ and ****, ‡‡‡‡ $p<0.0001$.

## Intracellular microscopic $D_{//}$, $D_{\perp}$ and µFA from metabolite DDES data

Representative metabolite DDES spectra from the PWM and OGM VOIs are shown as a function of θ in figures 5A and 5B. The resonances of tNAA, tCr and tCho could be identified and reliably quantified using LCModel in both regions. Cramér–Rao lower bounds (PWM/OGM) for the spectral fit (as percent of the standard deviation) were 4±1/4±1, 10±2/8±2 and 8±1/12±3 for tNAA, tCr, and tCho respectively. Values refer to measurements at $b=7199$ s/mm$^2$ and were averaged across subjects and values of θ. Panels C and D in figure 5 show the fitting results of the θ-modulated signal (averaged over subjects) for the PWM and the OGM VOIs. Figure 6 reports the averaged values for metabolite $D_{//}$, $D_{\perp}$, and µFA and the results of the statistical analysis. For all three metabolites, $D_{//}$ was significantly lower in OGM compared to PWM ($p \leq 0.024$). $D_{\perp}$ and µFA were not significantly different between the two VOIs, except for tCho ($p \leq 0.043$ for both $D_{\perp}$ and µFA). $D_{//}$(tCho) was significantly lower than $D_{//}$(tNAA) in both PWM and OGM ($p ≤ 0.003$), and µFA(tCho) was significantly lower compared to µFA(tNAA) only in OGM ($p = 0.030$). Finally, µFA > 0.8 for all three intracellular metabolites in both VOIs, except for tCho in OGM.

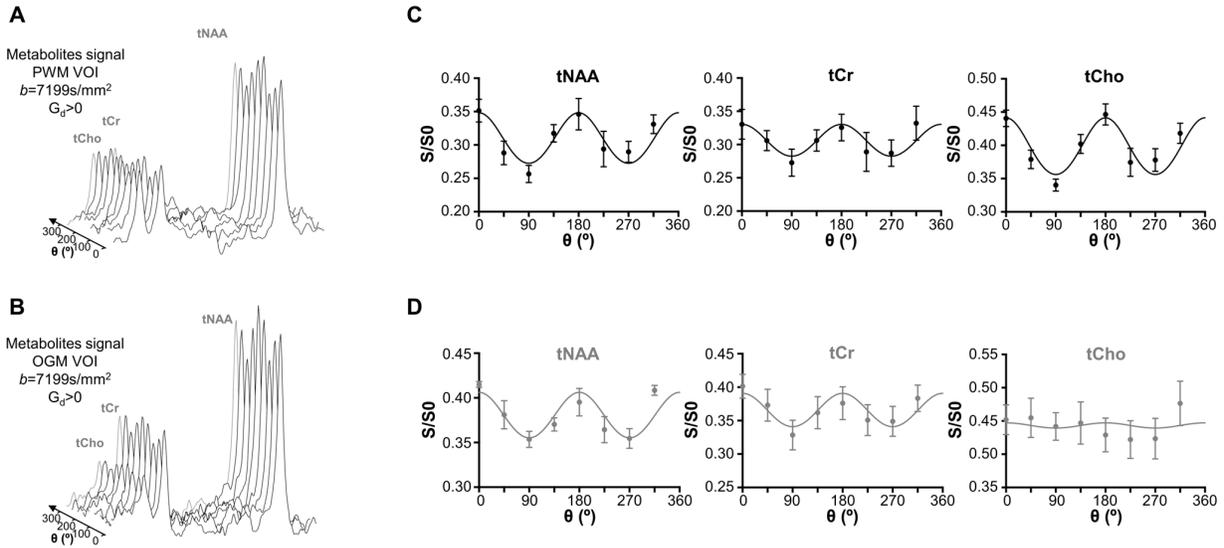

*Figure 5*: Illustration of individual DDES metabolite spectra for different θ in (A) PWM and (B) OGM VOIs. Following signal quantification, the θ-modulation for tNAA, tCr, and tCho in both (C) PWM (black) and (D) OGM (gray) VOIs were fitted using an ensemble of uniformly rotated axisymmetric diffusion tensors. The data (circle) and fits (solid line) for the mean over all participants are illustrated for both VOIs, and the error bars indicate the standard error of the mean.

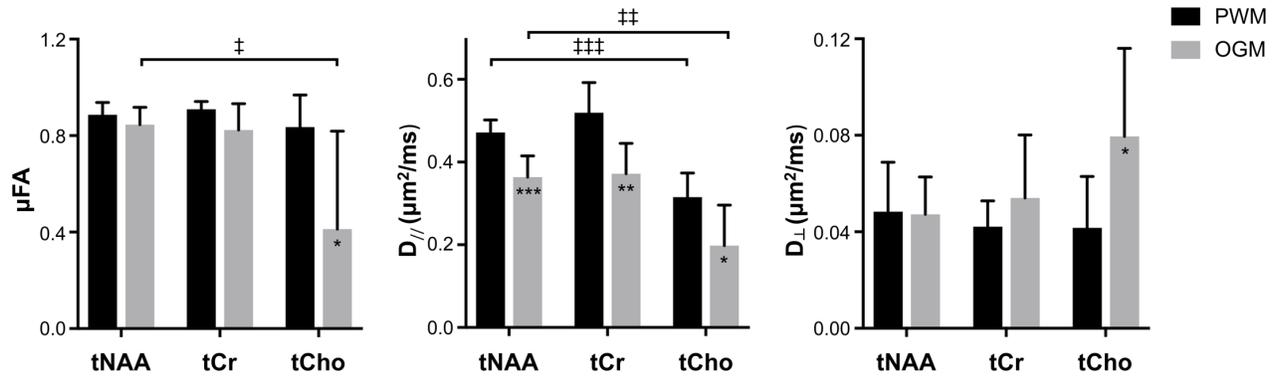

*Figure 6*: Fitted model parameters for the metabolites data (mean±s.d.). Statistical significance between PWM and (large) OGM (represented with *) and between tCho and tNAA (represented ‡) in both VOIs was evaluated using an unpaired Student's t-test with unequal variance. *, ‡ $p < 0.05$, **, ‡‡ $p < 0.005$ and ***, ‡‡‡ $p < 0.001$.

## $D_{//}$, $D_⊥$ and µFA from water DDES data over a range of *b* values

### *Water DDES at the highest b value*

The water signal was quantified over a range of *b* value in the PWM and OGM VOIs (figure 7). Table 3 reports the averaged values and statistical analyses for $S_0$, $D_{//}$, $D_⊥$, and µFA. $S_0$ was evaluated from each consecutive pair of *b* values and normalized to the $S_0$ at the lowest *b* (918 s/mm$^2$) as explained in the methods section, and is referred to as "fitted $S_0$" from here on. At the highest $b$, the fitted $S_0$ was significantly lower in OGM VOI compared to PWM VOI. In both VOIs, µFA(water) > 0.8 at the highest $b$, and significantly higher in the PWM VOI ($p$ = 0.0008). Differences in µFA(water) were associated with differences in $D_{//}$(water) as well as in $D_{\perp}$(water). $D_{//}$(water) was significantly higher in the PWM VOI compared to OGM VOI ($p$ < 0.0001). $D_{//}$(water) was also significantly higher than $D_{//}$(tNAA) in both PWM and OGM VOIs ($p$ < 0.0001). Finally, $D_{\perp}$(water) was significantly lower in PWM VOI compared to OGM VOI ($p$ < 0.0009). $D_{\perp}$(water) was significantly higher than $D_{\perp}$(tNAA) in OGM VOI ($p$ < 0.0003). Values for the sOGM VOI are reported in appendix (table S1), however due to the small sample size no statistical test were performed.

Figure 8 illustrates $D_{//}$(tNAA) and $D_{//}$(water) as a function of GM and WM content across subjects and brain regions. Tissue fractions were calculated with respect to the sum of GM and WM within the VOI, assuming that the contribution from CSF at the highest $b$ is fully suppressed. $D_{//}$(water) and $D_{//}$(tNAA) increase with the fraction of WM in the VOI ($R^2$=0.83 and 0.74 respectively, $p$ < 0.0003). Using reported values[8] for $D_{free}$(water) (3 µm$^2$/ms) and $D_{free}$(tNAA) (0.78 µm$^2$/ms) at 37°C, we estimated the intracellular tortuosity, T, for water and tNAA at 100% WM and 100% GM, where T = $\sqrt{D_{free}/D_{//}}$ (figure 8E, table 4).

For each tissue type (100% WM and 100% GM), similar tortuosity values for tNAA and water were obtained. Tortuosity values for OGM were significantly higher than those in PWM ($p$ < 0.003 for both water and tNAA). $D_{//}$ for tCr and tCho were also estimated, both indicating lower $D_{//}$ for these metabolites in GM than in WM.

*Water DDES across b values*

We observed a significantly lower µFA(water) at all $b$ < 7199 s/mm$^2$ in OGM VOI ($p$ <0.0008) when compared to the µFA(water) at the highest $b$ value(see table 3 and figure 9). µFA(water) remained relatively unchanged in PWM across $b$ values, with a significantly lower µFA(water) from $b$=2066 s/mm$^2$ ($p$ < 0.031). A significantly higher $D_{//}$(water) was observed at the two lowest $b$ values ($p$ < 0.0005) when compared to the values at the highest $b$ value in all VOIs. $D_{\perp}$(water) was also significantly higher at lower $b$ values (from $b$ ≤ 2060 s/mm$^2$ in PWM and $b$ ≤ 4050 s/mm$^2$ in OGM VOIs, $p$ < 0.02).

µFA(water) was significantly lower in OGM VOI compared to PWM VOI at all $b$ < 7199 s/mm$^2$ ($p$ < 0.0003) At the lowest $b$, µFA(water) was more than three times higher in PWM VOI compared to OGM VOI. $D_{//}$(water) in OGM VOI was significantly different than $D_{//}$(water) in the PWM VOI for all $b$ values ($p$ < 0.003). $D_{//}$(water) was significantly lower in OGM VOI compared to PWM VOI at $b$ > 2066 s/mm$^2$ and was higher at the lowest $b$. $D_{\perp}$(water) was significantly higher in OGM VOI compared to $D_{\perp}$(water) in the PWM VOI at $b$ < 7199 s/mm$^2$ ($p$ < 0.0001).

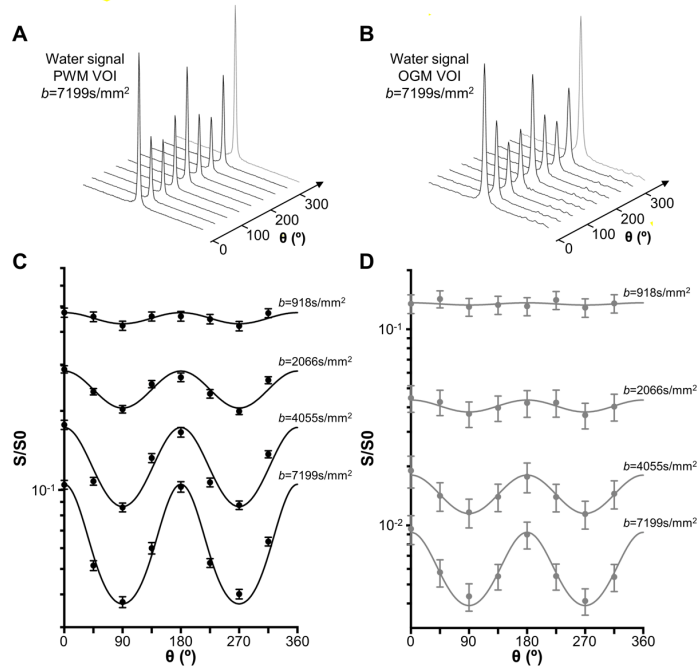

*Figure 7*: Individual water spectra (geometric mean of positive and negative gradient polarities) for different θ in (A) PWM and (B) OGM VOIs. Following signal quantification, the θ-modulation for both (C) PWM and (D) OGM VOIs was fitted using an ensemble of uniformly rotated axisymmetric diffusion tensors. The data (circles) and fits (solid line) for the mean over all participants are illustrated for both VOIs, error bars indicate the standard error.

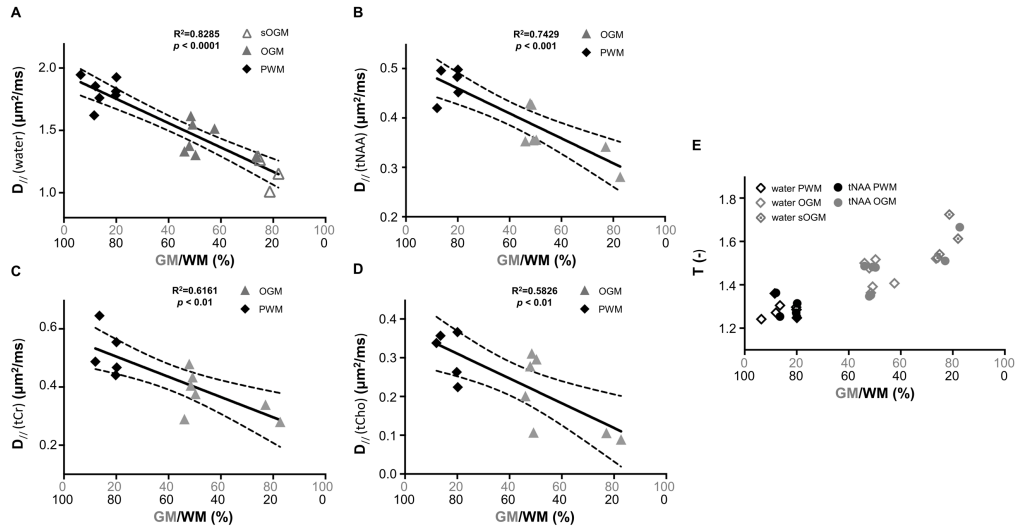

*Figure 8*: Correlation between GM/WM tissue fraction (%) and (A) water (B) tNAA (C) tCr, and (D) tCho parallel diffusivities ($D_{//}$). Data for different VOIs are represented with different markers: PWM VOI as filled black diamonds, OGM and sOGM VOIs as filled and open gray triangles, respectively. The solid line shows the linear regression and the dashed lines represent the 95% confidence intervals. (E) Tortuosity values are estimated for tNAA (filled black circle) and water (open diamond) as a function of GM/WM tissue fraction (%). The regression lines for the tortuosity are not displayed for clarity. $R^2 > 0.8$ and $p < 0.0002$ for both tNAA and water.

|  |  | $b$=918 s/mm$^2$ | | $b$=2066 s/mm$^2$ | | $b$=4050 s/mm$^2$ | | $b$=7199 s/mm$^2$ | |
|---|---|---|---|---|---|---|---|---|---|
| **Fitted S$_0$** | **PWM** | 1.00 | | 0.86±0.14 ‡ | **** | 0.74±0.09 | **** | 0.65±0.10 | **** |
| | **OGM** | 1.00 | | 0.38±0.12 ‡‡‡ | | 0.16±0.07 ‡ | | 0.08±0.05 | |
| **D$_{//}$** | **PWM** | 2.36±0.19 ‡‡‡ | ** | 2.07±0.08 ‡‡‡ | * | 1.94±0.08 | *** | 1.83±0.11 | **** |
| **(µm$^2$/ms)** | **OGM** | 2.90±0.34 ‡‡‡‡ | | 1.91±0.13 ‡‡‡‡ | | 1.56±0.12 | | 1.43±0.13 | |
| **D$_\perp$** | **PWM** | 0.28±0.15 ‡ | **** | 0.13±0.06 ‡ | **** | 0.08±0.03 | **** | 0.06±0.02 | *** |
| **(µm$^2$/ms)** | **OGM** | 1.93±0.34 ‡‡‡‡ | | 0.77±0.10 ‡‡‡‡ | | 0.32±0.06 ‡‡‡‡ | | 0.12±0.03 | |
| **µFA** | **PWM** | 0.87±0.07 ‡ | **** | 0.93±0.03 ‡ | **** | 0.96±0.01 | *** | 0.97±0.01 | *** |
| | **OGM** | 0.25±0.08 ‡‡‡‡ | | 0.52±0.07 ‡‡‡‡ | | 0.76±0.06 ‡‡‡ | | 0.91±0.03 | |

*Table 3*: *Fitted model parameters for the water data (mean ± s.d.). Statistical significance was evaluated using an unpaired Student's t-test with unequal variance. * denotes significance level in difference between PWM and OGM in the same table cell. ‡ denotes significance level in differences between the diffusion measure/VOI at the b value in that column and the same diffusion measure/VOI at the highest b value (b = 7199 s/mm$^2$). *, ‡ p<0.05, **, ‡‡ p<0.005, ***, ‡‡‡ p<0.001, ****, ‡‡‡‡ p<0.0001, and ns non significant.*

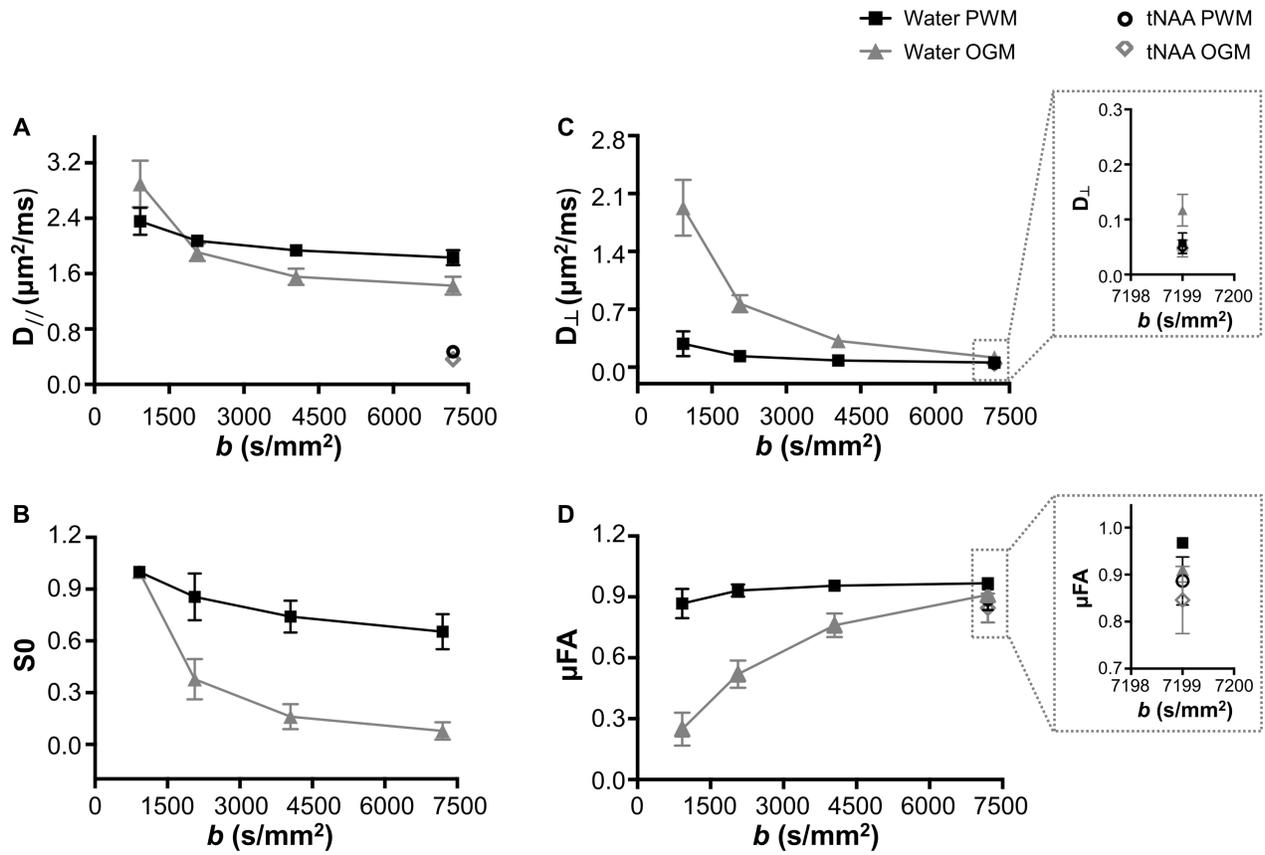

*Figure 9*: Water $D_{//}$ *(A)*, $D_{\perp}$ *(B)*, $S_0$ *(C) and µFA (D) as a function of b value for the three brain regions: PWM VOI (black solid line) and OGM VOI (gray solid line). $D_{\perp}$ and µFA values for tNAA in PWM (back circles) and OGM (gray diamonds) VOIs at the highest b value are also displayed in the zoomed boxes.*

# Discussion

In this work we demonstrate an approach for studying local microstructural features of brain tissue compartments by measuring and analyzing side-by-side water and metabolite DDES data. While metabolite DDES measurements provide an unequivocal empirical benchmark for intracellular diffusion metrics of neuronal and glial metabolites, water DDES measured with a range of - values enables the gradual elimination of the CSF and extracellular contributions, offering a reading of the intracellular diffusivity alone, as well as indirect information on the local geometry of the extracellular space. We used this approach to study the characteristics of intracellular diffusion processes of water and metabolites in GM and WM, highlighting common features as well as stark differences between the compartmental properties of these two tissue types. Table 4 summarizes the salient microstructural metrics obtained from the DDES measurements of metabolites and those of water at the highest $b$, highlighting our findings that pertain to the intracellular space in both GM and WM.

### *Intracellular spaces are highly anisotropic but different in GM and WM*

With the exception of µFA(tCho) in OGM, the µFA of all three metabolites was higher than 0.8 in both OGM and PWM. At the highest $b$, where it is assumed that faster-diffusing extracellular and CSF contributions are suppressed, µFA(water) was also very high (~0.9) and comparable with µFA(tNAA) and µFA(tCr) in both regions, as well as with µFA(tCho) in white matter. This strongly suggests that in the time scale of our measurements, the intracellular space is highly anisotropic in both GM and WM.

This is evident in WM, where the high µFA for both intracellular water and the three metabolites suggests a predominant neurite morphology for all cell types. Our observation of a highly anisotropic intracellular space in WM supports the hypothesis of the prevalence of thin fibers in WM, which could consist of long-range axons or processes of fibrous astrocytes. The high µFA of the preferentially glial metabolite tCho suggests that the characteristic distance between branching points in WM glia exceeds the diffusion length in our DDES experiments, and within the diffusion/mixing time of our experiments tCho experiences a mostly simple "single fiber" environment. This is supported by the typical cytomorphology of fibrous astrocytes in WM, characterized by long, thin, and relatively unbranched structures in contrast to their protoplasmic counterparts in GM[47,48]. The notion of a predominant intracellular contribution from neurites in both GM and WM in the human brain and other mammals has been expressed in previous reports, both from the DW-MRS literature[9,10,19,49,50] as well as more recently from the DWI literature[51]. The high µFA values for intracellular water, tNAA, and tCr in GM support the notion that neurites dominate as a microstructural unit also in GM.

The high µFA in GM is in stark contrast to the overall low *macroscopic* anisotropy in GM obtained from DTI measurements, which reflects the overall lack of directionality in neurite propagation across the measurement volume, even at the millimeter scale. This has been alluded to in studies that investigated microscopic anisotropy in post-mortem tissue[35,36,52,53] as well as from DDE imaging studies in the human brain performed at low *b* values[38], and is strong supported by this current study, with the added value of cross-validation between water and metabolite data. Noteworthy is that the µFA(water) values obtained in GM in a previous report[38] are lower than those found in WM and lower than those reported here. At the *b* value in which these experiments were performed (total diffusion weighting of $b < 1000$ s/mm$^2$) it is likely that signal from the more isotropic extracellular space in GM significantly contributed to the overall signal.

The only exception to high intracellular µFA in GM in our measurements was the µFA(tCho). This can be interpreted in two possible ways. One is the lower tCho signal to noise ratio (SNR) for the single measurement, resulting in higher variability of the tCho signal across gradient orientations in the DDES results. The other possibility is that in GM, a significant fraction of the predominantly glial tCho is found in protoplasmic astrocytes. These astrocytes, found extensively in human GM, are highly branched cells, significantly more so than their fibrous counterparts in WM[48]. It is plausible that the effective µFA(tCho) is low because the average diffusion length of tCho is comparable or higher than the distance between branching points on processes in protoplasmic astrocytes. Additional support for this explanation is also provided by the lower $D_{//}$ of tCho in GM compared to the one in WM, mentioned in the following paragraph. The uniqueness of diffusion properties of tCho in human GM has been previously reported[40], where the sub diffusion index α for tCho in GM was significantly lower than that of tCr and tNAA in GM and all three metabolites in WM. In figure 10 are shown schematic representations of microstructural features such as deviation from propagation along a straight line and branching processes, that may affect microscopic diffusion metrics at different diffusion lengths.

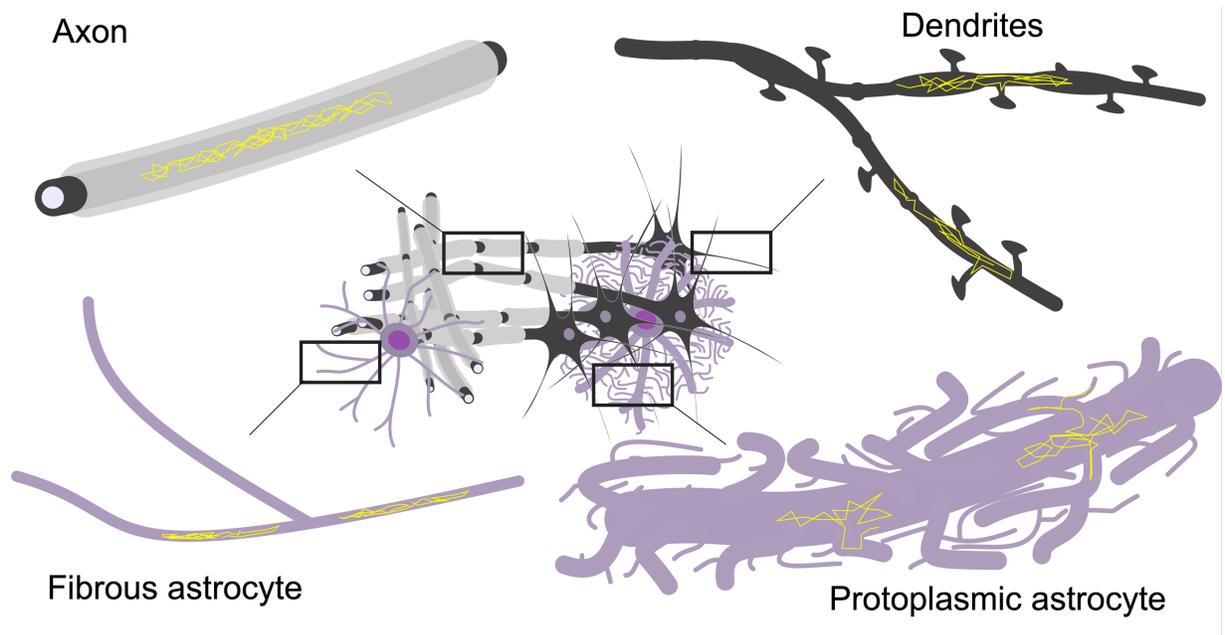

*Figure 10*: Schematic representation of microstructural features such as deviation from propagation along a straight line and branching processes, that may affect microscopic diffusion metrics in neurons, fibrous, and protoplasmic astrocytes. Axons, dendrites and fibrous astrocytes are thin fibrous structures with few branches which may locally constitute highly anisotropic diffusion. Protoplasmic astrocytes are however distinctly different with highly branched and thicker processes which could provide a more isotropic diffusion.

***Diffusion properties of the intracellular space are different in GM and WM***

Correlations with tissue fraction and direct comparisons between data from OGM and PWM both indicate that the intracellular parallel diffusivity $D_{//}$ obtained from metabolite DDES data was significantly different across tissue types as well as across metabolites within the same tissue type. $D_{//}$ for all three metabolites was lower in OGM than in PWM, indicating an overall lower diffusivity in the intracellular space in GM. In good correspondence with this finding, $D_{//}$ of water evaluated from the highest *b* value was lower in OGM than in PWM, and similarly to the metabolite data, $D_{//}$ of water at the highest *b* value was strongly negatively correlated with GM fraction in the VOI. This tight correspondence between the tissue dependence of $D_{//}$ of metabolites and $D_{//}$ of water at the highest *b* value further corroborates our hypothesis that the extracellular water is effectively suppressed at $b = 7199$ s/mm$^2$ and that the diffusional properties of water at this high *b* value reflect almost exclusively those of the intracellular water.

We used the strong correlation between $D_{//}$ of both water and metabolites at the highest *b* value with tissue type fraction to estimate intracellular $D_{//}$ of water and the three metabolites in pure GM and WM. Assuming that within the diffusion encoding times (~45 ms) in our experiments the diffusion length along the neurite represents a path along a straight-propagating unbranched fiber (< 15 µm), $D_{//}$ can be seen as the cytoplasmic diffusion coefficient for the metabolites and the intracellular water. When branches and deviation from straight propagation occur within the diffusion length at a given $t_d$, these geometric features will influence $D_{//}$ as well (see figure 10). We observed a lower GM $D_{//}$ for all metabolites, as well as for water at high *b*. Differences in $D_{//}$ of tNAA between GM and WM may reflect differences in mitochondrial density between WM axonal fibers and GM neurons[54] as well as morphological differences between long propagating WM axons and highly branched dendritic trees in cortical neurons. Differences in $D_{//}$ of tCr, and mostly of tCho between the two regions may also reflect cytomorphological differences between astrocytes in both regions, as mentioned in the previous paragraph regarding µFA(tCho). Differences in metabolite $D_{//}$ across tissue types are also consistent with tissue-specific metabolite ADC measurements obtained in previous DW-MRS studies[40,45,55–58].

Based on the $D_{//}$ values of water and tNAA and their free diffusion coefficients at 37ºC, we estimated the intracellular tortuosity experienced by water and tNAA, where the tortuosity of tNAA reflects exclusively that inside neurons. As vividly shown in figure 8E, the tortuosity values obtained for water and tNAA are remarkably similar within the same tissue type despite a 3-4 fold difference between the $D_{//}$ for water and tNAA, and vary significantly between GM and WM. The intracellular tortuosity reflects all extrinsic factors that may slow down the mobility of molecules in the intracellular space, such as differences in viscosity, geometrical obstacles in the cytosol such as organelles or undulations, and varicosities (variation in diameter along the fiber) of the fibrous structures. For example, the similarity between the tortuosity of water and tNAA in both tissue types suggests that despite the differences in their free diffusivity, a similar coarse-grained environment is probed by both molecules over the diffusion encoding period. Similar relations have previously been shown for a number of metabolites and ions in macroscopic ADC measurements in rodents[59]. This excellent agreement between two independent measurements suggests that the combined diffusion measurements of both water and metabolites can provide a valuable platform for validating models of water diffusion also on a microscopic scale in both GM and WM. We focused on the calculation of tortuosity of tNAA (T(tNAA)) because of the specificity to one particular cell type (neurons), as well as because of the relatively small contribution of the co-measured metabolite (NAAG) to the results. The tortuosity of tCr and tCho is more difficult to estimate without accurate information on the fractions of co-measured metabolites in tCr (Cr and PCr) and tCho (Cho, GPC and PCho), each with its specific molecular weight and concentration. Similarly, the values for free diffusivity of tCr and tCho require better assessment in phantom experiments with a realistic combination of co-measured metabolites that represents these combinations *in vivo* in GM and WM. We intend to pursue the investigation of the unique diffusion properties of tCr and tCho more thoroughly in future studies.

Values for $D_{//}$(water) in WM we report here align well with earlier studies using different approaches for filtering extracellular and CSF contributions[12,29,30]. Similar values have also been found in more model-driven analyses of conventional DWI data, which also suggest two-fold differences in axonal and dendritic intracellular diffusivities[28]. Estimating all contributions to the water signal from different tissue compartments provides a rather flat fitting landscape for selecting the right combination of fractions and diffusivities[28,60], making the fitting procedure for compartment-specific microscopic diffusion metrics fairly unstable. The spatial resolution of conventional DWI is in general on the order of the cortical thickness, making direct measurements without contaminations from CSF and WM all but impossible. In our study, we operated on an even coarser resolution, many times over the cortical thickness. We demonstrated that taking into account tissue fractions within the VOI, it is possible to obtained consistent and reliable microstructural details on GM via correlation analyses. This can be easily extended to imaging studies where partial volume within DWI voxels can be obtained in a similar way to the one we used here[61].

$D_{//}$ across metabolites within the same tissue type were also significantly different from one another. These differences were already observed in single diffusion encoding (SDE) DW-MRS experiments in humans and animal models[9,40,45,49,56,62–65]. Interpretation of these values will have to take into account differences in cytoplasm across cell types, the potential effect of cytomorphological features, and molecular weight differences across the metabolites.

$D_{\perp}$(tNAA) in WM we report here is higher than the one we reported in a previous work, where we estimated the microscopic diffusion metrics of tNAA based on analysis of powder-averaged data[10]. There, reported values for $D_{\perp}$(tNAA) were between 0.011 and 0.024 µm$^2$/ms. Underestimation and larger variance of µFA, equivalent to an underestimation of $D_{//}$ and overestimation of $D_{\perp}$, may be attributed to a limited SNR[10,66]. Our current measurements were conducted at longer TE, and with less averages compared to our previous study (ref. 10), leading to a lower SNR in our present study. We estimated the effect of SNR on the estimation of $D_{//}$ and $D_{\perp}$ of tNAA with a simulation that takes into account our experimental settings (figure 11). We assumed that $D_{\perp}$ = 0 (displacement RMS perpendicular to the fiber wall >> fiber diameter) and $D_{//}$ = 0.5 µm$^2$/ms which are representable values for tNAA[19,67]. SNR is defined here as the tNAA signal relative to the standard deviation of the noise in the *b* = 0 condition. In our experiments,

SNR for the tNAA peak estimated by LCModel was in the range of 30-45 for the data averaged across DW conditions. The simulation shown here indicates that $D_\perp$ of tNAA would be overestimated to about 0.02-0.03 µm$^2$/ms, with a standard deviation on the same order. These estimates are slightly lower but in the same range of our $D_\perp$(tNAA) as estimated from our current data. A similar low $D_\perp$(water) was estimated in PWM but with a 2-3 fold increase in the more GM-rich OGM and sOGM voxels. We speculate that this may reflect residual extracellular signal, either from non-complete filtering or comparably higher transmembrane exchange in the unmyelinated dendrites. Additional intracellular contributions from protoplasmic astrocytes contributing to the lower µFA(tCho) discussed above may also play a role. It is expected, however, that the effect of a nonfinite axonal radius on $D_\perp$ is small[68] within the range of diffusion times and diffusion coefficients in our experiments.

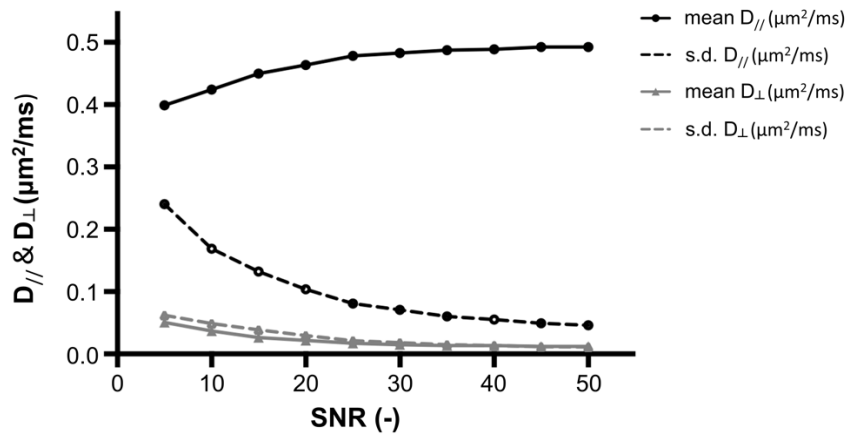

*Figure 11*: Noise propagation in the parameter estimates for settings comparable to the metabolite acquisition for a substrate with $D_{//}$ = 0.5 µm$^2$/ms and $D_\perp$ = 0 µm$^2$/ms ("stick" situation) comparable to tNAA in white matter. Mean and standard deviations are estimated from 1000 realizations of the data with Gaussian noise at different SNR. Metabolite data presented in this paper has an SNR range ~30-45.

|  |  |  | PWM | OGM | WM | GM |
|---|---|---|---|---|---|---|
|  | GM (%) |  | 15.4±5.3 | 56.3±14.5 | 0 | 100 |
|  | WM (%) |  | 84.6±5.3 | 40.8±14.5 | 100 | 0 |
| From modeling individual DDES data | $D_{//}$ (µm²/ms) | Water | 1.83±0.11 | 1.43±0.13 |  |  |
|  |  | tNAA | 0.47±0.03 | 0.36±0.05 |  |  |
|  | $D_{\perp}$ (µm²/ms) | Water | 0.06±0.02 | 0.12±0.03 |  |  |
|  |  | tNAA | 0.05±0.02 | 0.05±0.02 |  |  |
|  | µFA | Water | 0.97±0.01 | 0.91±0.03 |  |  |
|  |  | tNAA | 0.89±0.05 | 0.85±0.07 |  |  |
|  | T | Water | 1.28±0.04 | 1.45±0.07 |  |  |
|  |  | tNAA | 1.29±0.04 | 1.47±0.10 |  |  |
| From $D_{//}$ and tissue fraction correlation | $D_{//}$ (µm²/ms) | Water | 1.80±0.12 | 1.40±0.26 | 1.95±0.06 | 0.97±0.17 |
|  |  | tNAA | 0.47±0.04 | 0.41±0.08 | 0.51±0.02 | 0.25±0.07 |
|  | T | Water | 1.29±0.04 | 1.46±0.14 | 1.24±0.02 | 1.76±0.15 |
|  |  | tNAA | 1.29±0.06 | 1.38±0.14 | 1.24±0.03 | 1.74±0.23 |

*Table 4: Summary of all parameters extracted from the data acquired in PWM and OGM VOIs at highest b value (mean±s.d.).*

### *Morphology of the extracellular space varies significantly between WM and GM*

A significantly lower µFA in both PWM and OGM VOIs was observed at lower *b* values. Under the assumption that the contribution of the extracellular space to the measured signal decreases with increasing *b* value, this suggests that the extracellular space is less microscopically anisotropic than the intracellular space in both WM and GM, but that there is a significant degree of microscopic anisotropy in the extracellular space in WM, absent in GM. At the lowest *b*, water µFA is more than three times lower in OGM than in the PWM VOI, suggesting that the extracellular space in WM is highly anisotropic, or conversely, that the tortuosity in the direction perpendicular to the fiber direction significantly affects water diffusion in WM. This result is consistent with the fact that long axons are the dominant structural feature in WM and are relatively densely packed, resulting not only in a high intracellular anisotropy but also in a significant extracellular anisotropy.

**Data quality and limitations**

With DDES now established in human settings, there are some experimental conditions which could be further optimized in future studies. Data here were averaged over three orthogonal planes and collected from relatively large VOIs. These conditions should approach the powder average condition, which assumes full orientational dispersion of fibers within the volume. Some residual order, however, could affect the observed signal modulations. Moreover, while intermediate angles carry information regarding oblate and prolate shapes[52], the maximum contrast for microscopic anisotropy is expected between the parallel and perpendicular conditions, as shown in figure 3, panels C and 3D. With this in mind, optimized sampling schemes for powder averaging with only the parallel and perpendicular conditions could be considered, such as the 5-design proposed by Jespersen[13], or the minimal protocol by Yang[69]. The two protocols have recently been evaluated side-by-side with comparable performance[66].

Our data fitting approach differs slightly from the original definition of μFA based on contrasting $b^2$-terms (kurtosis) at low $b$ values reflecting variance in mean displacements across either directions or domains[13–15]. While this provides a stringent theoretical description, larger $b$ values might be needed to provide sufficient contrast-to-noise (see figure 3D), which in turn leads to the increasing influence of higher-order terms[53]. Here we followed an approach valid at all $b$, but under the assumption of mono-disperse gaussian domains, which was also applied in previous DDES studies[19,20]. Given the numerous potential interpretations of DDE/DDES data, a model-free representation of the data is given in the appendix (tables S2 and S3).

Finally, it is likely that the microstructural and diffusional properties in gray and white matter will vary across the brain and will be dictated by local myelo- and cytoarchitecture. This calls for expanding this type of investigations to other brain regions, or to incorporate DDES experiments in spatially resolved techniques such as magnetic resonance spectroscopic imaging (MRSI).

**Possible time-dependence effects**

While this study focused on the effects of microscopic anisotropy, the DDE experiment may entangle different temporal fingerprints of time-dependent diffusion phenomena, such as exchange across different domains[70] or reflections from restrictive barriers[71]. These effects could bias measurements of microscopic anisotropy but could also provide rich additional information regarding microstructure or physiological processes. Our measurement was performed at a relatively short and fixed mixing time (~5.3 ms) which will neither modulate nor capture exchange on longer time scales, e.g. between branches of highly arborized cells such as protoplasmic astrocytes in GM or transmembrane exchange in non-myelinated fibers.

Time-dependent effects from restrictions could however potentially generate a difference between the signal intensity at parallel (θ=0º) and anti-parallel (θ=180º) directions with a lower signal in the latter[71,72]. We investigated the size of this contrast (see figure S3 in the appendix), but found no significant difference for the metabolites, in contrast to results reported in a recent DDES study in rodents. Higher SNR or shorter and stronger gradient pulses in the animal setting could reflect a different spatial scale and explain this difference. We did find a significant difference between S(θ=0º) and S(θ=180º) for water in the lowest $b$ values in all VOIs. This suggests that water data over that range of $b$ values are time-dependent. Time-dependence could be explained by numerous effects in the extra- and intracellular environments, such as spatial disorder in axonal packing or variation in intracellular geometries[24]. The lack of effects in metabolites and water at high $b$ values may suggest that the effect probed by the time scale of the experiment is mainly extra-cellular[24,73].

Measurements over a range of $b$ values, mixing and diffusion times could indeed provide a more detailed view of the morphology of restrictions and the aforementioned exchange effects. Tuning the diffusion experiment to the expected phenomena with e.g. optimized gradient waveforms could be a promising approach for detecting independent effects that affect microscopic anisotropy and other local diffusion metrics[74,75]. Such data could also challenge more intricate biophysical models featuring multiple geometrical features, such as spherical cell bodies or connected and arborizing fibrous structures of particular relevance for the study of neuronal structures in GM or glial cells in general[76,77].

## Conclusion

We presented here a comprehensive approach for combining water and metabolite DDES to investigate the local microstructural and cytoplasmic features of extra- and intracellular spaces in the human brain. We demonstrated the usefulness of this approach by shedding light on differences as well as similarities in local anisotropy and cytoplasmic properties in different cell types in gray and white matter, and inferred on differences in local geometry between the two tissue types. This approach can be extended to combine other microstructural probes, such as *b* tensor encodings, and be further expanded to investigate time dependent effects on intra- and extracellular diffusion properties to further characterize brain tissue microstructure.

## Appendix

**Multi-compartmental effects**

A simple model can serve as a hypothetical model of the contributions of intracellular, extracellular, and CSF components of the signal at different *b* values. Simulated signals are shown in figure S1. Here, the intracellular diffusivity is modeled as a "stick" with zero axial diffusivity, the extracellular space as an anisotropic tensor, and CSF as an isotropic tensor. Examples of representative diffusivities are taken from figure 4 in Novikov et al[1] and for WM were set to : intracellular $D_{//}/D_\perp$ = 2/0 µm$^2$/ms, extracellular $D_{//}/D_\perp$ = 2/0.5 µm$^2$/ms; and for GM set to: intracellular $D_{//}/D_\perp$ = 1/0 µm$^2$/ms, extracellular $D_{//}/D_\perp$ = 0.75/0.5 µm$^2$/ms. CSF diffusivity was set to 3 µm$^2$/ms.

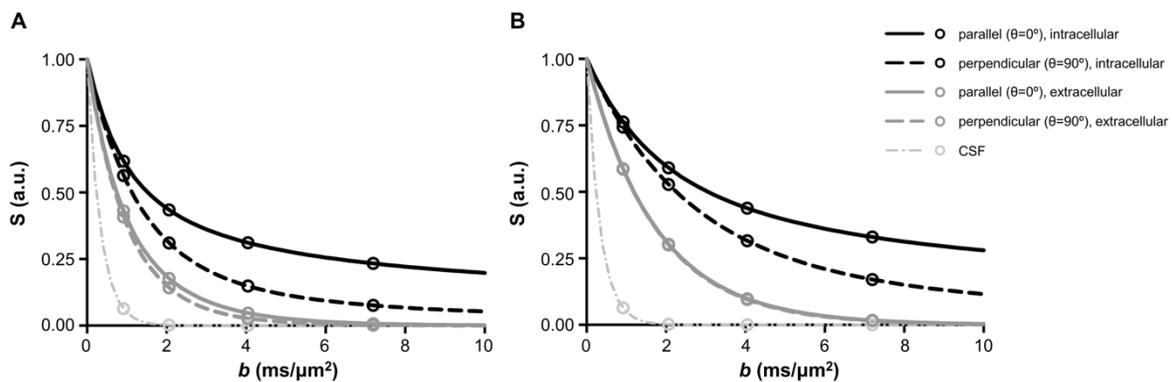

*Figure S1*: Simulated powder averaged signals for the parallel (solid lines) and perpendicular (dashed lines) conditions for three different signal components in WM (A) and GM (B). All signals are normalized. The b values used in the water acquisitions are shown with markers. At the highest b value, the intracellular component dominates and in all but the lowest b value, the CSF contribution is suppressed below <0.25%.

**Effect and correction of contaminating gradient fields - phantom data and simulations**

Cross-terms between the diffusion gradients and static background gradients or imaging gradients (such as the crushers and slice-selection gradients) may bias the measurement of $D_{//}$, $D_\perp$ and µFA. To investigate this effect we acquired all data with both positive and negative diffusion gradient polarities. The geometric mean of the signal over both conditions was then calculated which cancels out cross-terms to first order in *b* value. Our sequence and approach were validated *in vitro* using the "BRAINO" phantom (GE Medical Systems, Milwaukee, WI, USA). Water DW-spectra were acquired with two *b* values (0 and 1111 s/mm$^2$) and both diffusion gradient polarities. For each condition, the water signal was calculated as the peak integral. As illustrated in figure S2A, the behavior of the water signal as a function of θ changes with gradient polarities and as expected becomes independent of θ when taking the geometric mean of the signal acquired with both polarities. This demonstrates that the effects of cross-terms between diffusion and imaging gradients are then corrected. We further validated this using Matlab simulations calculating the signals from *b* tensors with and without contributions from background or imaging gradients. Diffusion and imaging gradients were extracted from a simulation of the sequence on the scanner software as described earlier. This showed that all cross-terms between diffusion and imaging gradients are canceled as well as cross-terms with spatially constant background gradients (figure S2B and C). However, the effects spatially varying microscopic background gradients on the length scale of the diffusion pathway are not cancelled.

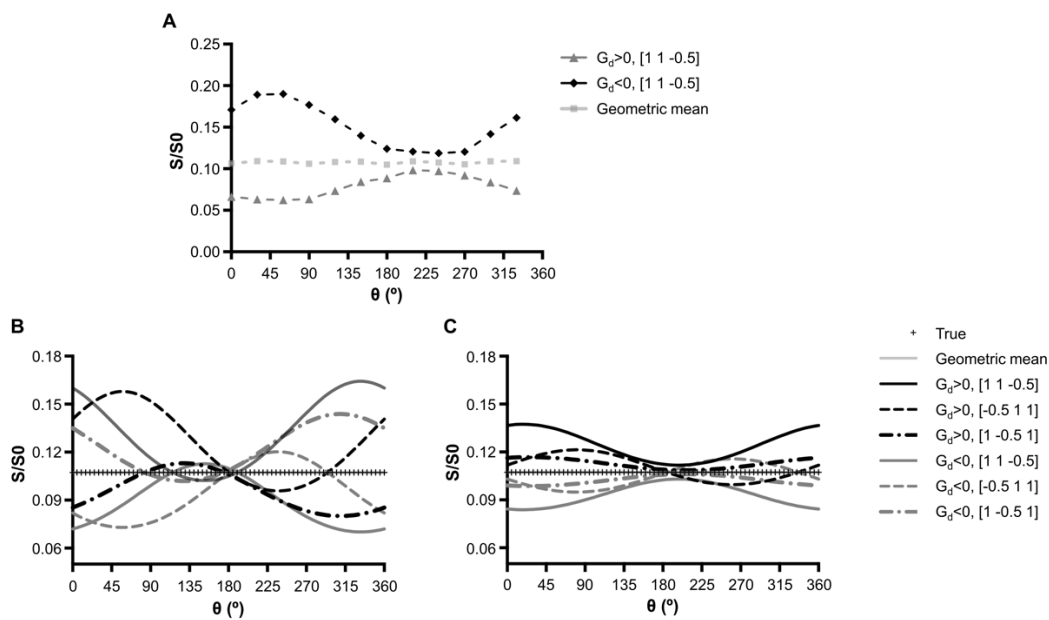

*Figure S2*: (A) Illustration of the effect of cross-terms between the diffusion gradients and imaging gradients in BRAINO (GE Medical Systems, Milwaukee, WI, USA) phantom is illustrated. This was further investigated using simulations showing the effect of cross-terms between (A) diffusion gradients and imaging gradients or (C) diffusion gradients and background gradients ($G_{background}$=[1 1 0] and 10mT/m).

**Time dependence**

If the mixing time in a DDE experiment is short compared to the characteristic length of a restriction a difference in signal between the anti-parallel and parallel conditions will be observed[2,3]. Diffusion in the proximity of barriers is likely to change direction from boundary reflections, leading to a larger signal attenuation when the encoding direction change is in the anti-parallel condition. In contrast, the anti-parallel condition is velocity compensated which means that ballistic intravoxel incoherent motion (IVIM) from e.g. disperse blood flow in capillaries, will be rephased leading to the opposite pattern[4,5]. Pairwise comparison of the two conditions is shown in figure S3. We observe significantly lower anti-parallel signals in the order of ~1% signal difference in the lower b values of the water acquisition in both PWM and OCC which support apparent effects from restrictions/reflections rather than IVIM effects.

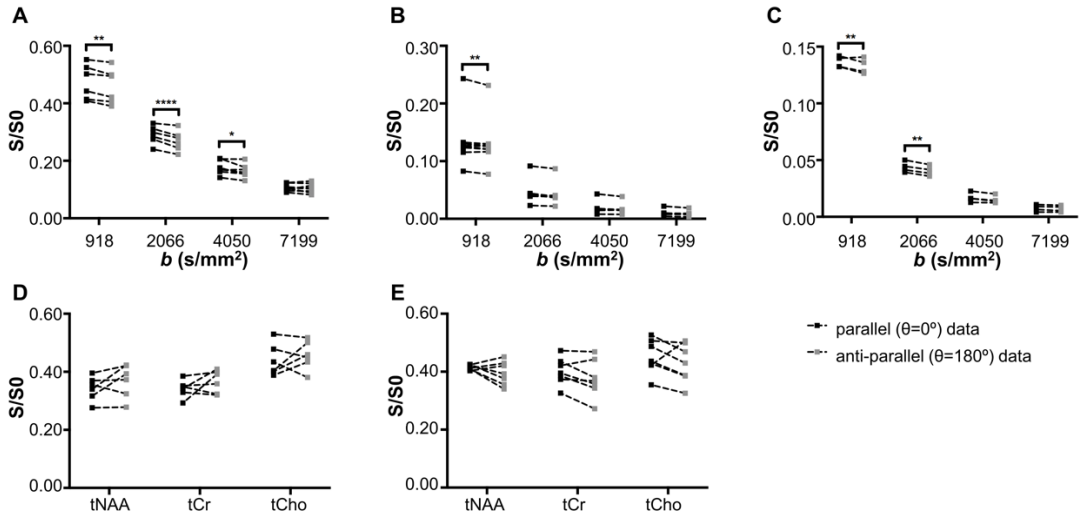

*Figure S3*: *Comparison of the difference in signal quantified with parallel (θ=0º) and anti-parallel (θ=180º) directions for water in (A) PWM, (B) OGM and (C) OGMs VOIs as well as for metabolites data in (D) PWM and (E) OGM VOIs at different b values. Connected lines (dashed lines) between data points (square) connect data from each participant's data. Significant differences between parallel and anti-parallel (represented with \*) lower b values were observed in both VOIs for water data. \* $p<0.05$, \*\* $p<0.005$ and \*\*\* $p<0.001$.*

**Small OGM VOI data**

Additional water data from 4 participants with a smaller VOI was collected in OGM to achieve a larger relative fraction of gray matter. The data is included in figure 8 in the main paper. Tissue volume fractions and fitted model parameters are shown in table S1.

| | | | | |
|---|---|---|---|---|
| **Gray Matter (GM, %)** | | | 67.7±4.5 | |
| **White Matter (WM, %)** | | | 19.8±3.9 | |
| **Cerebrospinal Fluid (CSF, %)** | | | 12.5±6.4 | |
| | $b$=918 s/mm$^2$ | $b$=2066 s/mm$^2$ | $b$=4050 s/mm$^2$ | $b$=7199 s/mm$^2$ |
| **Fitted S$_0$** | 1.00 | 0.38±0.03 | 0.15±0.01 | 0.07±0.02 |
| **D$_{//}$ (µm$^2$/ms)** | 2.91±0.15 | 1.85±0.15 | 1.37±0.13 | 1.18±0.13 |
| **D$_\perp$ (µm$^2$/ms)** | 1.84±0.10 | 0.79±0.09 | 0.35±0.09 | 0.18±0.08 |
| **µFA** | 0.28±0.06 | 0.49±0.06 | 0.69±0.10 | 0.82±0.10 |

*Table S1*: Volume fraction (%, mean±s.d.) of WM, GM, and CSF and fitted model parameters for the water data (mean±s.d.) in the sOGM VOI.

**Simple representation of signal**

A simple representation of the signal can serve as an input for alternative interpretations. The relevant information (disregarding possible effects from time dependence) is the offset and amplitude of the angular modulation. In the tables S2 and S3 we state the values from fitting the following equation to the normalized data as suggested in earlier studies[6,7]:

$$S(\theta) = A + B \cdot \cos 2\theta$$

| | | tNAA | tCr | tCho |
|---|---|---|---|---|
| **A** | **PWM** | 0.310±0.047 | 0.297±0.045 | 0.406±0.068 |
| | **OGM** | 0.380±0.019 | 0.365±0.053 | 0.443±0.063 |
| **B** | **PWM** | 0.038±0.011 | 0.042±0.011 | 0.024±0.009 |
| | **OGM** | 0.026±0.008 | 0.025±0.012 | 0.004±0.016 |

*Table S2*: Phenomenological representation of the signal offset A and modulation amplitude B for metabolites at $b$ = 7199 s/mm$^2$.

|   |      | $b$=918 s/mm$^2$ | $b$=2066 s/mm$^2$ | $b$=4050 s/mm$^2$ | $b$=7199 s/mm$^2$ |
|---|------|------------------|-------------------|-------------------|-------------------|
|   | PWM  | 0.452±0.058 | 0.243±0.024 | 0.125±0.014 | 0.064±0.008 |
| A | OGM  | 0.135±0.045 | 0.041±0.020 | 0.015±0.008 | 0.006±0.003 |
|   | sOGM | *0.435±0.055* | *0.230±0.026* | *0.120±0.018* | *0.061±0.007* |
|   | PWM  | 0.022±0.005 | 0.039±0.006 | 0.043±0.006 | 0.033±0.005 |
| B | OGM  | 0.002±0.001 | 0.003±0.002 | 0.003±0.002 | 0.003±0.001 |
|   | sOGM | *0.019±0.003* | *0.035±0.006* | *0.040±0.007* | *0.030±0.002* |

*Table S3*: Phenomenological representation of the signal offset A and modulation amplitude B for water at different b values.